\documentclass[letterpaper]{article} 
\usepackage{aaai25}  
\usepackage{times}  
\usepackage{helvet}  
\usepackage{courier}  
\usepackage[hyphens]{url}  
\usepackage{graphicx} 
\usepackage{natbib}  
\usepackage{caption} 
\usepackage{algorithm}
\usepackage{algorithmic}
\usepackage{amssymb} 
\usepackage{amsmath} 
\usepackage{booktabs}
\usepackage{multirow}
\usepackage{graphicx}
\usepackage{subcaption}

\usepackage{newfloat}
\usepackage{listings}
\DeclareCaptionStyle{ruled}{labelfont=normalfont,labelsep=colon,strut=off} 
\floatstyle{ruled}
\newfloat{listing}{tb}{lst}{}
\floatname{listing}{Listing}
\title{Disentangled Modeling of Preferences and Social Influence \\for Group Recommendation}
\author{
    Guangze Ye\textsuperscript{\rm 1, \rm 2, \rm 3},  
    Wen Wu\textsuperscript{\rm 3, \rm 4}\thanks{The corresponding author.},
    Guoqing Wang\textsuperscript{\rm 3},
    Xi Chen\textsuperscript{\rm 4},
    Hong Zheng\textsuperscript{\rm 5},
    Liang He\textsuperscript{\rm 1, \rm 2, \rm 3}
}
\affiliations{
    \textsuperscript{\rm 1}Lab of Artificial Intelligence for Education, East China Normal University, Shanghai, China\\
    \textsuperscript{\rm 2}Shanghai Institute of Artificial Intelligence for Education, East China Normal University, Shanghai, China\\
    \textsuperscript{\rm 3}School of Computer Science and Technology, East China Normal University, Shanghai, China\\
    \textsuperscript{\rm 4}Shanghai Key Laboratory of Mental Health and Psychological Crisis Intervention, School of Psychology and Cognitive Science, East China Normal University, Shanghai, China\\
    \textsuperscript{\rm 5}Shanghai Changning Mental Health Center, Shanghai, China\\

    \{gzye, wgq\}@stu.ecnu.edu.cn, wwu@cc.ecnu.edu.cn, xchen@psy.ecnu.edu.cn, zhhmm2@163.com, lhe@cs.ecnu.edu.cn
}

\usepackage{bibentry}

\begin{document}

\maketitle

\begin{abstract}
The group recommendation (GR) aims to suggest items for a group of users in social networks. Existing work typically considers individual preferences as the sole factor in aggregating group preferences. Actually, social influence is also an important factor in modeling users' contributions to the final group decision. However, existing methods either neglect the social influence of individual members or bundle preferences and social influence together as a unified representation. As a result, these models emphasize the preferences of the majority within the group rather than the actual interaction items, which we refer to as the preference bias issue in GR. Moreover, the self-supervised learning (SSL) strategies they designed to address the issue of group data sparsity fail to account for users' contextual social weights when regulating group representations, leading to suboptimal results. To tackle these issues, we propose a novel model based on \underline{\textbf{Dis}}entangled Modeling of Preferences and Social Influence for Group \underline{\textbf{Rec}}ommendation (DisRec). Concretely, we first design a user-level disentangling network to disentangle the preferences and social influence of group members with separate embedding propagation schemes based on (hyper)graph convolution networks. We then introduce a social-based contrastive learning strategy, selectively excluding user nodes based on their social importance to enhance group representations and alleviate the group-level data sparsity issue. The experimental results demonstrate that our model significantly outperforms state-of-the-art methods on two real-world datasets.
\end{abstract}

%

\begin{figure}[t]
\centering
\includegraphics[width=1\columnwidth]{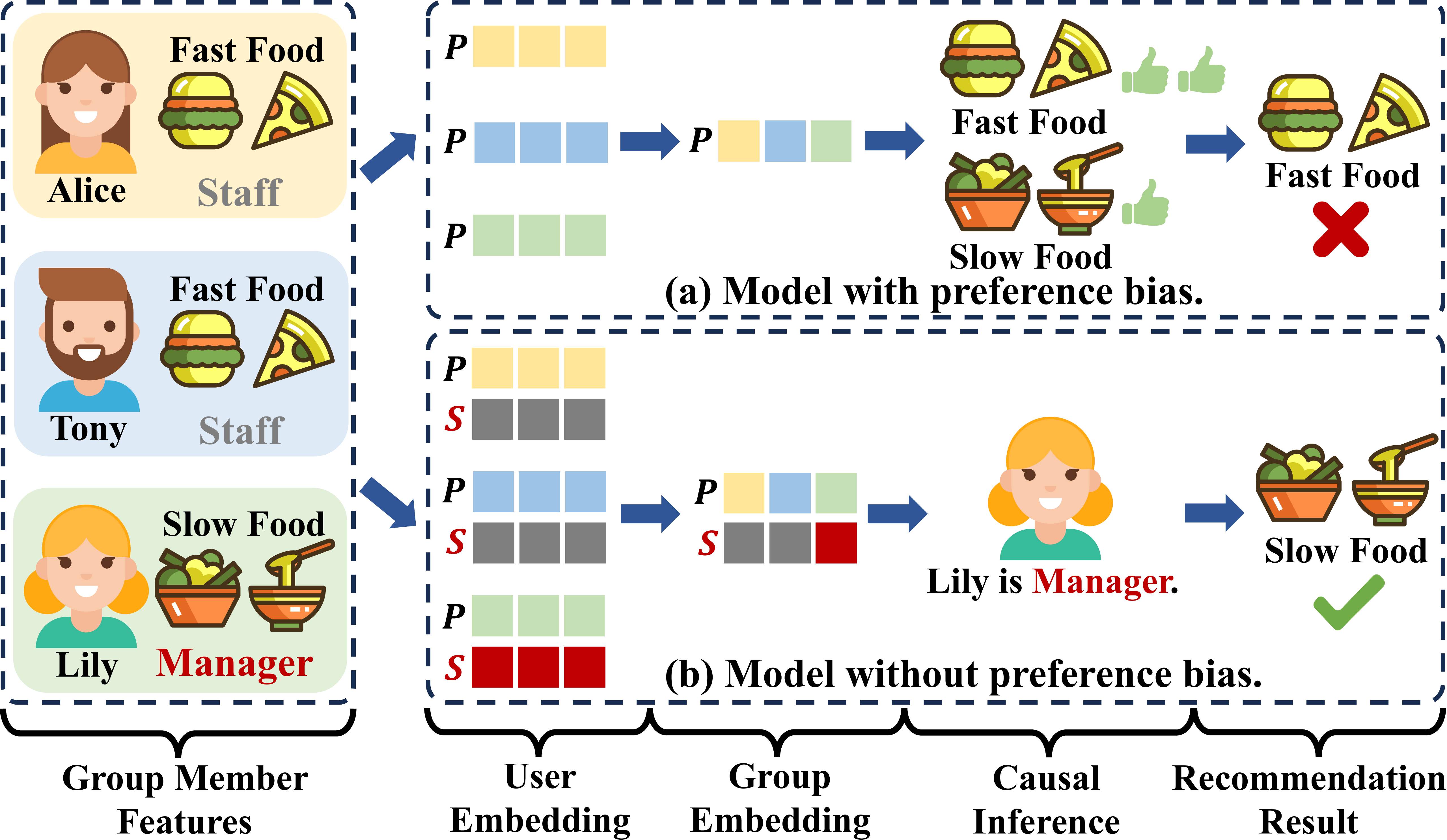} 
\caption{(a) and (b) show the causal inference process of the model with and without preference bias, where $P$ and $S$ are preferences and social embedding, respectively.}
\label{fig1}
\end{figure}  

\section{Introduction}
Humans, as social animals, have a strong desire to participate in group activities. Consequently, a task called GR has emerged \citep{hu2014deep}. The GR aims to simulate the group decision-making process and suggest items for a group of users in social networks \citep{xiao2017fairness}. In existing works \citep{cao2018attentive, sankar2020groupim, wang2020group, zhang2021double,jia2021hypergraph, wu2023consrec, Li2023SelfSupervisedGG, ye2024personality}, GR generally consists of three steps: (1) Training individual preferences; (2) Aggregating individual preferences to obtain group preferences; and (3) Optimizing group preferences. For example, AGREE \citep{cao2018attentive} aggregates member preferences through attention mechanisms, while CubeRec \citep{chen2022thinking} utilizes hypercubes to aggregate user representations. Although these methods have achieved satisfactory results, they still suffer from the following two limitations.  

On one hand, these methods overly focus on individual preference factors, thereby neglecting other important factors affecting group recommendations, such as social influence, which helps quantify and differentiate the contributions of group members to a group decision \cite{zheng2018identifying, yin2019social}. Although some studies \citep{yin2019social, guo2020group} consider social information, they fail to disentangle user preferences and social influence but rather bundle them together as a unified representation. Such entanglement leads the model to emphasize the preferences of the majority within the group, which we refer to as the preference bias issue in GR. We provide an example in Figure 1 to further illustrate this bias. Alice and Tony, who are colleagues, prefer fast food, while Lily, their manager, prefers slow food. In (a), the model fails to disentangle preferences and social influence, leading to incorrect recommendations of fast food favored by the majority in the group. In (b), the model assigns group members with separate embedding for preferences and social influence, avoiding preference bias issues. Consequently, the model correctly infers that the group's choice of slow food is due to Lily's greater influence as a manager. Disentangled user representations thus help the model accurately determine each member's contribution to group decisions, eliminating preference bias in GR. We illustrate the causal relationships of members' contributions through Figure 2.  

\begin{figure}[t]
\centering
\includegraphics[width=0.45\columnwidth]{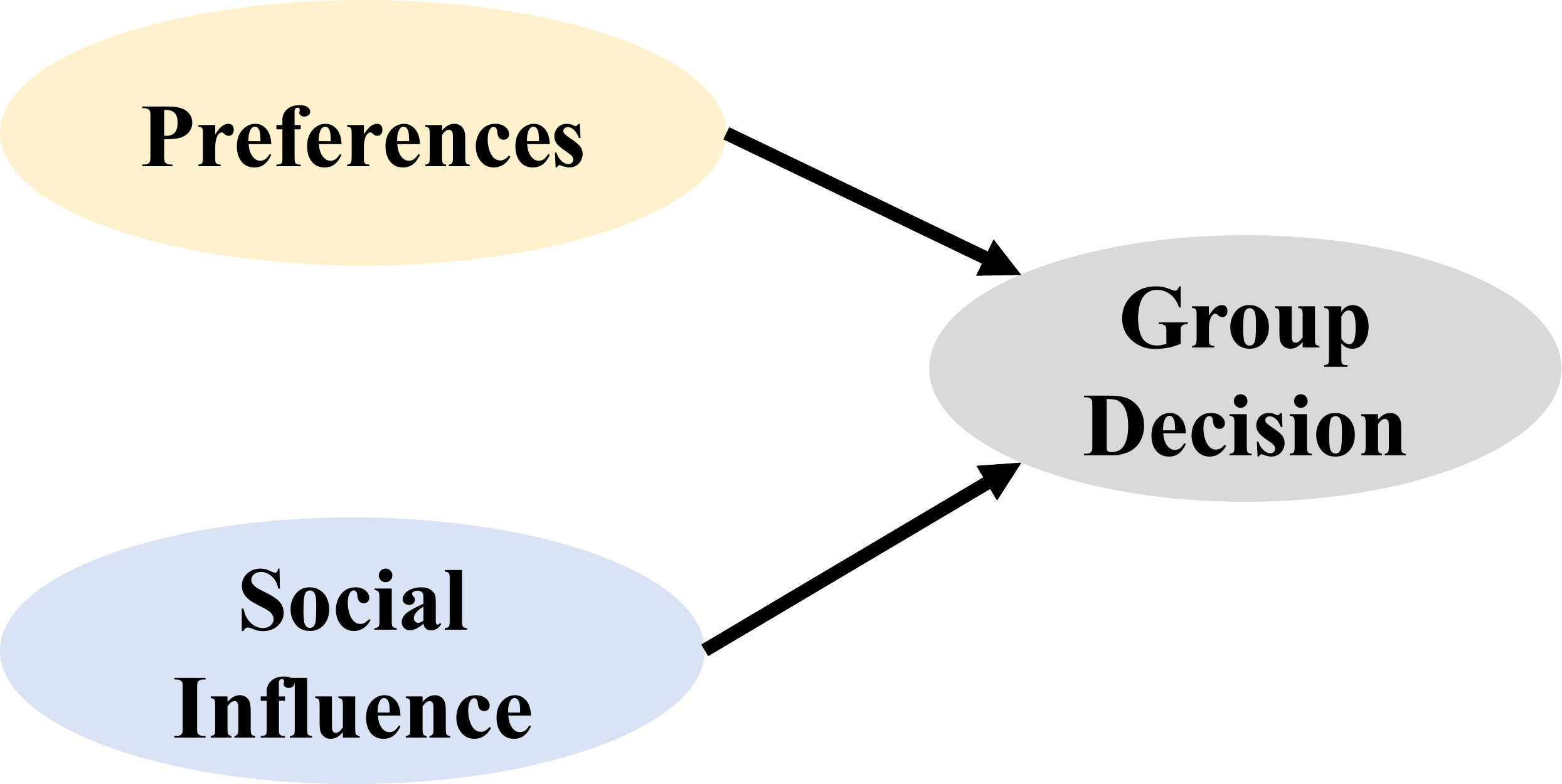}
\caption{The casual graph of GR.}
\label{fig2}
\end{figure}    

On the other hand, most methods suffer from the group-item interaction data sparsity issue \citep{sankar2020groupim}. Although a few methods alleviate this issue by designing different SSL objectives \citep{sankar2020groupim, zhang2021double, Li2023SelfSupervisedGG}, these SSL strategies fail to account for users' contextual social weights when regulating group representations, resulting in suboptimal performance. For example, HHGR \cite{zhang2021double} and SGGCF \cite{Li2023SelfSupervisedGG} achieve data augmentation by randomly removing member nodes from the group and aligning the group representations before and after removal. Since members contribute differently to the group's final decisions, randomly removing member nodes can alter the decision-making process. In this scenario, such an alignment strategy is harmful.

To address these challenges, we propose a novel model based on \underline{\textbf{Dis}}entangled Modeling of Preferences and Social Influence for Group \underline{\textbf{Rec}}ommendation (DisRec). Firstly, 
we design an innovative user-level disentangling network to tackle the preference bias issue in GR. Specifically, we start by devising a disentangled embedding layer to separately encode user preferences and social influence, with two individual graph structures developed to capture their respective features. To comprehensively capture users' social influence, we innovatively integrate the user-user social graph with the user-item hypergraph \cite{bai2021hypergraph, feng2019hypergraph} into a unified social hypergraph, employing an attention convolution mechanism for embedding propagation. Next, we concatenate the disentangled embedding to form final representations for users and items, deriving user-level group representations through an attention aggregator. 

Secondly, we construct a group co-occurrence graph to capture the inherent preferences of the group independently of individual preferences, inspired by previous works \citep{cao2018attentive, jia2021hypergraph, wu2023consrec}. We then use a gate aggregator to merge user-level and group-level representations into the final group representations. 

Thirdly, we introduce a novel social-based contrastive learning strategy to enhance group representations and alleviate the group-level data sparsity issue. Unlike previous methods that randomly remove user nodes \citep{zhang2021double, Li2023SelfSupervisedGG}, we selectively exclude user nodes based on their contextual social importance. Our key technical insight is that only highly influential users significantly affect group decisions. Thus, we remove the most influential user to maximize the divergence between the resultant and original group representations, and the least influential user to minimize this divergence.

To summarize, the contributions are listed as follows: 

\begin{itemize}
\item We are the first to identify the issue of preference bias in GR and to formulate a causal graph for GR, where the group's final decision stems from two independent factors, namely user preferences and social influence. 
\item We propose a novel method, DisRec, which addresses the issue of preference bias in GR by disentangling user representations. Besides, it introduces a social-based SSL strategy that alleviates group-level data sparsity.
\item We conduct extensive experiments on two real-world datasets to verify the superiority of DisRec over state-of-the-art baselines. The code for DisRec is available at (https://github.com/DisRec/DisRec).
\end{itemize}

\section{Related Work}
\subsection{Group Recommendation}

As mentioned before, existing GR methods first learned member preferences from U-I interaction data, then aggregated these member preferences into group preferences, and finally optimized the group preferences using group-item interaction data. Early methods used predefined aggregation strategies, such as average \cite{rakesh2016probabilistic}, least misery \cite{baltrunas2010group}, and maximum pleasure \cite{boratto2011state}. Later works \citep{cao2018attentive, vinh2019interact, he2020game, yin2019social, guo2020group, Deng2021KnowledgeAwareGR, zhang2022gbert,  ji2023multi} adopted attention networks to explicitly model the importance of group members. Recent research employed more complex structures to achieve better user and group representations, such as hypercubes \citep{chen2022thinking} and hypergraphs \citep{guo2021hierarchical, zhang2021double, jia2021hypergraph, wu2023consrec}. However, these methods either ignored the social influence of individual members or used a unified representation for users, thereby failing to disentangle user preferences and social influence. Furthermore, other methods attempted to design SSL strategies to address the sparsity of group data in GR \citep{zhang2021double, sankar2020groupim, Li2023SelfSupervisedGG, chen2022thinking}. For example, SGGCF \citep{Li2023SelfSupervisedGG} created different views by randomly dropping out user nodes and edges, which cannot guarantee the quality of generated positive and negative examples. 

\subsection{Disentangled Representation Learning in RS} 
There were some recent studies \citep{zheng2021disentangling, wang2022disentangled, li2022disentangled, wu2022disentangled, chen2023bias, fang2024backdoor} that utilized disentangled representation learning to alleviate bias in recommendation systems. For example, DICE \citep{zheng2021disentangling} addressed popularity bias by disentangling user interests and conformity, while DISGCN \citep{li2022disentangled} focused on separating social homophily and influence in social recommendations. In this work, we disentangle users' preferences and social influence to eliminate preference bias in GR. 

\section{Preliminaries}
Let $\mathcal{U}=\left\{u_1, u_2, \ldots, u_M\right\}$, $\mathcal{I}=\left\{i_1, i_2, \ldots, i_N\right\}$ and $\mathcal{G}=\left\{g_1, g_2, \ldots, g_K\right\}$ to represent the sets of $M$ users, $N$ items and $K$ groups, respectively. There are three kinds of observed interaction data among $\mathcal{U}$, $\mathcal{I}$ and $\mathcal{G}$: user-item interactions, group-item interactions and user-user interactions. We use $\mathbf{Y}_U \in \mathbb{R}^{M \times N}$ to denote the user-item interactions, where the element $\mathbf{Y}_U(i, j) = 1$ if user $u_i$ has interacted with item $i_j$ otherwise $\mathbf{Y}_U(i, j) = 0$. Likewise, we use $\mathbf{Y}_G \in \mathbb{R}^{K \times N}$ and $\mathbf{R} \in \mathbb{R}^{M \times M}$ to denote the group-item interactions and user-user interactions, respectively. The $t$-th group $g_t \in\mathcal{G}$ consists of a set of users, and we use $\mathcal{G}_t = \left\{u_1, u_2, \ldots, u_s, \ldots, u_{|\mathcal{G}_t|}\right\}$ to denote the user set of $g_t$, where $u_s \in \mathcal{U}$, and ${|\mathcal{G}_t|}$ is the size of group $g_t$. We denote the set of items which are interacted by $g_t$ as $\mathcal{Y}_t = \left\{i_1, i_2, \ldots, i_j, \ldots, i_{|\mathcal{Y}_t|}\right\}$, where $i_j \in \mathcal{I}$, ${|\mathcal{Y}_t|}$ is the size of $\mathcal{Y}_t$. Then, given a target group $g_t$ (or target user $u_j$) , our goal is to generate a ranked list of items that $g_t$ (or target user $u_j$) may be interested in.

\textbf{Definition 1. Hypergraph.} Let $\mathcal{G}^h = (\mathcal{V}^h, \mathcal{E}^h)$ denote a hypergraph, where $\mathcal{V}^h$ is the vertex set and $\mathcal{E}^h$ is the hyperedge set. Each hyperedge $e \in \mathcal{E}^h$ contains two or more vertices and is assigned a positive weight $\boldsymbol{w}_e$. $\mathbf{W} \in \mathbb{R}^{|\mathcal{E}^h| \times |\mathcal{E}^h|}$ is the diagonal matrix of the weight of the hyperedge. The hypergraph can be represented by an incidence matrix $\mathbf{H} \in \mathbb{R}^{|\mathcal{V}^h| \times |\mathcal{E}^h|}$, where $\boldsymbol{h}_{i,e} =1$ if the hyperedge $e \in \mathcal{E}^h$ contains a vertex $v_i \in \mathcal{V}^h$, otherwise 0. For each hypergraph, $\mathbf{D}$ is the diagonal degree matrices of vertex, and $\mathbf{B}$ is the diagonal degree matrices of hyperedge. 

\textbf{Definition 2. Social Hypergraph.} On the basis of Definition 1, we further define the social hypergraph. The social hypergraph is a heterogeneous graph that includes two sets of edges: hyperedges and social edges. Let $\mathcal{G}^s = (\mathcal{V}^s, \mathcal{E}^h, \mathcal{E}^s)$ denote a social hypergraph, where
$\mathcal{E}^s=\left\{e_s=\left\{v_i, v_j\right\} \mid v_i, v_j \in \mathcal{V}^s, v_i \neq v_j\right\}$ is the social edge set. 

\textbf{Definition 3. Co-occurrence Graph.} The co-occurrence graph uses hyperedges from a hypergraph as nodes. If two hyperedges share at least one common node, there is an edge between them in the co-occurrence graph. Formally, given a hypergraph $\mathcal{G}^h = (\mathcal{V}^h, \mathcal{E}^h)$, a co-occurrence graph is denoted as $\mathcal{G}^g = (\mathcal{V}^g, \mathcal{E}^g)$ where $\mathcal{V}^g=\left\{e \mid e \in \mathcal{E}^h \right\}$ and $\mathcal{E}^g=\left\{\left(e_p, e_q\right) \mid e_p, e_q \in \mathcal{E}^h,\left|e_p \cap e_q\right| \geq 1\right\}$. Each edge is assigned a weight $\boldsymbol{w}_{p, q} = \frac {\left|e_p \cap e_q\right|}{\left|e_p \cup e_q\right|}$, and the adjacency matrix is defined as $\mathbf{A}\in \mathbb{R}^{M \times M}$ where $\boldsymbol{a}_{p,q}=\boldsymbol{w}_{p, q}$.

\begin{figure*}[t]
\centering
\includegraphics[width=0.89\textwidth]{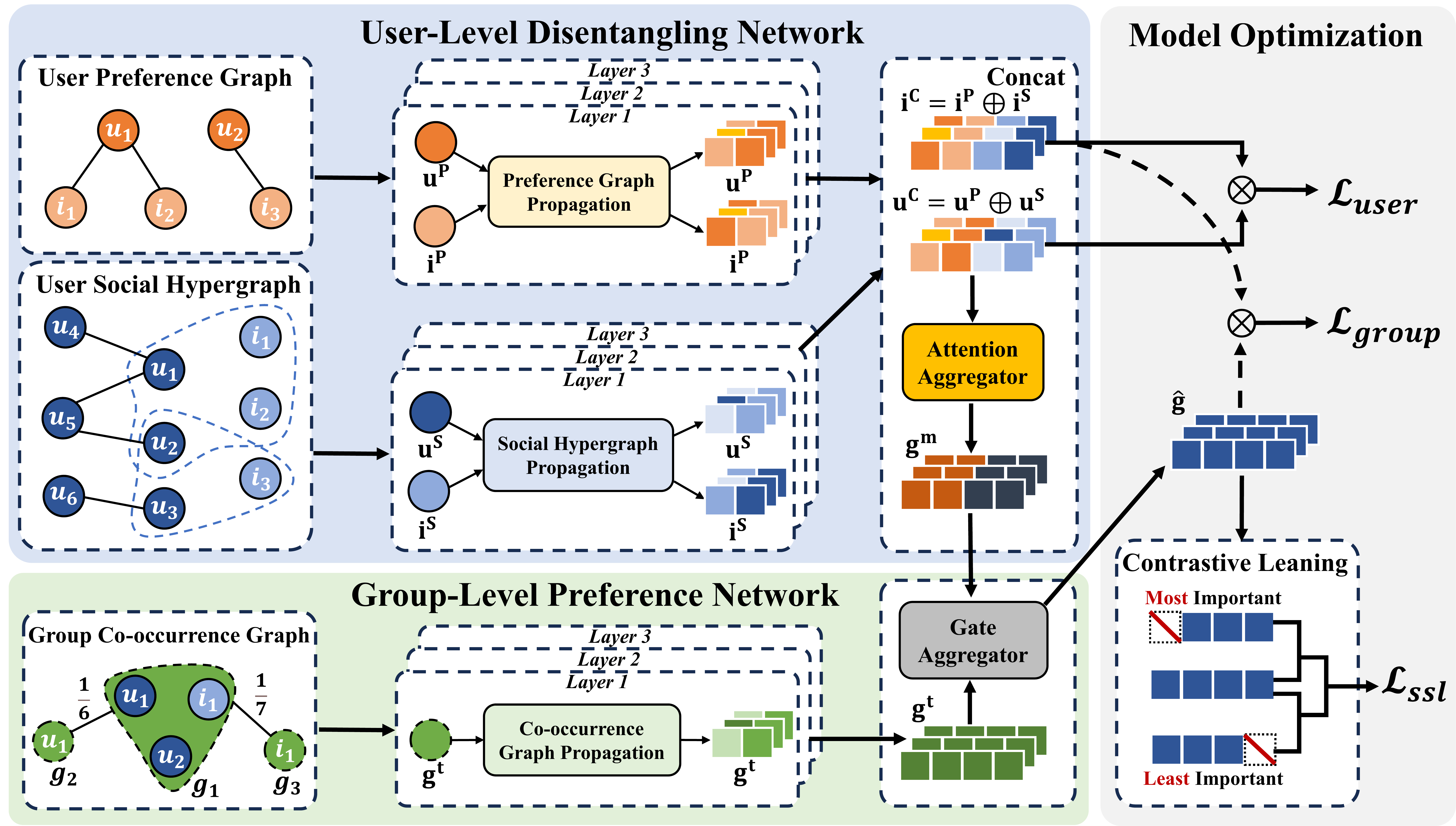} 
\caption{Framework of Disentangled Modeling of Preferences and Social Influence for Group Recommendation (DisRec).}
\label{fig3}
\end{figure*}  

\section{Proposed Model}

\subsection{User-Level Disentangleing Network}
\subsubsection{Graph Construction.} To disentangle the modeling of group members' preferences and social influence, we extract two types of graphs from U-I and G-I interactions: the user preference graph and the user social hypergraph, as shown on the left side of Figure 3. Specifically, the construction details of the two graphs are as follows:

\begin{itemize}
\item \textbf{User Preference Graph:} To capture each user's pure preferences, we construct the user preference graph $\mathcal{G}^p = (\mathcal{V}^p, \mathcal{E}^p)$, where $\mathcal{V}^p = \mathcal{U}$ represents the node set of users, and $\mathcal{E}^p = \{(u_t, i_j)|u_t \in \mathcal{U}, i_j \in \mathcal{I}, \mathbf{Y}_U(t,j)=1\}$ denotes the edge set comprising interactions between users and items. The adjacency matrix $\hat{\mathbf{A} }\in \mathbb{R}^{(M+N) \times (M+N)}$ represents the structure of $\mathcal{G}^p$.
\item \textbf{User Social Hypergraph:} 
To disentangle the social influence of members, we construct a social hypergraph $\mathcal{G}^s = (\mathcal{V}^s, \mathcal{E}^h, \mathcal{E}^s)$ where $\mathcal{V}^s = \mathcal{U}$ denotes the node
set of users, $\mathcal{E}^h = \mathcal{G}$ denotes the hyperedge set, and $\mathcal{E}^s=\left\{e_s=\left\{v_i, v_j\right\} \mid v_i, v_j \in \mathcal{U}, v_i \neq v_j\right\}$ denotes the social edge set. For user $u_i$ in group $g_t$, they are connected to other group members and interacted items via hyperedge $\mathcal{E}^h_t = \mathcal{G}_t \cup \mathcal{Y}_t$, and to their social friends via social edges $\mathcal{E}^s_{u_i}$. 
\end{itemize}

\subsubsection{Disentangled Embedding Layer.} 
Unlike existing work, we model two distinct factors using two sets of embeddings for users and items. Specifically, we use two embedding matrices, $\mathbf{U^P} \in \mathbb{R}^{d \times |\mathcal{U}|}$ and $\mathbf{U^S} \in \mathbb{R}^{d \times |\mathcal{U}|}$, to respectively represent user preferences and social influence, where $d$ is the embedding size. Similarly, to ensure consistency in the latent space, we assign embeddings for items that correspond to those of users: $\mathbf{I^P} \in \mathbb{R}^{d \times |\mathcal{I}|}$ and $\mathbf{I^S} \in \mathbb{R}^{d \times |\mathcal{I}|}$.

\subsubsection{Preference Graph Propagation.}
We utilize graph neural networks \citep{he2020lightgcn} for embedding propagation to represent the pure preferences of group members for items by capturing the collaborative signals between users and items. Specifically, we feed the concatenation of user preference embeddings $\mathbf{U^P} \in \mathbb{R}^{d \times |\mathcal{U}|}$ and item preference embeddings $\mathbf{I^P} \in \mathbb{R}^{d \times |\mathcal{I}|}$ to the graph convolutional network, denoted as $\mathbf{X_P^{(0)}}=[\mathbf{U^P};\mathbf{I^P}]$. Referring to the spectral graph convolution \citep{he2020lightgcn}, we define our graph convolution in the $l$-th layer as: 
\begin{equation} 
\mathbf{X_P}^{(l+1)}=\mathbf{\hat{D}^{-\frac{1}{2}}}\hat{\mathbf{A}}\mathbf{\hat{D}^{-\frac{1}{2}}}\mathbf{X_P}^{(l)},
\end{equation}
where $\mathbf{X_P}^{(l)}$ denotes the node representations in the $l$-th layer and $\mathbf{\hat{D}}$ is the diagonal node degree matrix of adjacency matrix $\mathbf{\hat{A}}$. After passing $\mathbf{X_P}^{(0)}$ through $L$ convolutional layers, we average the embeddings obtained at each layer to get the final representation $\mathbf{\hat{X}_P}=\frac{1}{L+1}\sum^L_{l=0}\mathbf{X_P}^{(l)}=[\mathbf{\hat{U}^P};\mathbf{\hat{I}^P}]$. Here, $\mathbf{\hat{U}^P}$ encodes information about the types of items that users will be interested in, while $\mathbf{\hat{I}^P}$ encodes information about the types of users that will like the items.

\subsubsection{Social Hypergraph Propagation.}
The ability of hyperedges to model group representations is widely considered in existing methods \citep{zhang2021double, wu2023consrec}. However, these methods are limited to using hyperedges only to represent group preferences, overlooking the social influence contained within hyperedges. Our method extracts social influence within groups from hyperedges and outside groups from the social graph. As for items, they can also fully leverage hyperedge representations to capture item features. 

Therefore, we first apply the hypergraph convolution operation to extract users' social influence within groups and items' social features.  Specifically, we feed the concatenation of user preference embeddings $\mathbf{U^S} \in \mathbb{R}^{d \times |\mathcal{U}|}$ and item preference embeddings $\mathbf{I^S} \in \mathbb{R}^{d \times |\mathcal{I}|}$ to the graph convolutional network, denoted as $\mathbf{X_S^{(0)}}=[\mathbf{U^S};\mathbf{I^S}]$. Referring to the spectral hypergraph convolution \citep{feng2019hypergraph, wu2019simplifying}, we define our hypergraph convolution as: 

\begin{equation} 
\mathbf{X_S}^{(l+1)}=\mathbf{D^{-1}}\mathbf{HWB^{-1}}\mathbf{H}^T\mathbf{X_S}^{(l)}\mathbf{\Psi}^{(l)},
\end{equation}
where $\mathbf{X_S}^{(l)}$ is the node representations in the $l$-th layer and $\mathbf{\Psi} \in \mathbb{R}^{d \times d}$ is the learnable weight matrix between two convolutional layers. In this process, the multiplication of $\mathbf{H}^T$ and $\mathbf{X_S}^{(l)}$ extracts the social information of each group, while the multiplication with $\mathbf{H}$ propagates this social information to each individual, forming their social influence within groups. We then perform a weighted sum of user friends' social influence within groups to generate users' social influence outside the group:

\begin{equation}  
\mathbf{u}_{\mathbf{S},i}^{(l+1)}=\mathbf{u}_{\mathbf{S},i}^{(l)}+\sum_{j \in \mathcal{N}(i)} \alpha_{j}\mathbf{W}_{1}\mathbf{u}_{\mathbf{S},j}^{(l)},
\end{equation} 
\begin{equation}
\alpha_{j}=\frac{exp(\mathbf{h}_1^T \mathbf{W}_{1} \mathbf{u}_{\mathbf{S},j}^{(l)})}{\sum_{{j'} \in\mathcal{N}(i)} exp(\mathbf{h}_1^T \mathbf{W}_{1} \mathbf{u}_{\mathbf{S},j'}^{(l)})},
\end{equation}
where $\mathbf{u}_{\mathbf{S},i}^{(l)} \in \mathbf{U^S}$ denotes the node representation of user $u_i$ in the $l$-th layer, $\alpha_{j}$ indicates the importance of the social relation from user $u_j$ to $u_i$, $\mathbf{W}_{1}$ and $\mathbf{h}_1^T$ are learnable parameters, and $\mathcal{N}(i)$ is the social friends of user $u_i$. Finally, we average the embedding obtained at each layer to generate the final representation $\mathbf{\hat{X}_S}=\frac{1}{L+1}\sum^L_{l=0}\mathbf{X_S}^{(l)}=[\mathbf{\hat{U}^S};\mathbf{\hat{I}^S}]$. Here, $\mathbf{\hat{U}^S}$ can be considered as a representation of how users will influence others in group decision-making process, while $\mathbf{\hat{I}^S}$ signifies the influence associated with items.

\subsubsection{User-Level Representation Aggregation.} 
We concatenate the thoroughly disentangled representations of preferences and social influences to obtain the final representations of users and items:
$
\mathbf{\hat{U}} = \mathbf{\hat{U}^P} || \mathbf{\hat{U}^S},
\mathbf{\hat{I}} = \mathbf{\hat{I}^P} || \mathbf{\hat{I}^S}, 
$
where $||$ means concatenation of two embeddings.  

Considering the varying contributions of individual users to the final group decision-making, we employ an attention aggregator to weightedly aggregate members’ representations: 
\begin{equation}  
\mathbf{g}^m_t=\sum_{j \in \mathcal{G}_t} \beta_{j}\mathbf{W}_{2}\hat{\mathbf{u}}_{j},
\end{equation} 
\begin{equation}
\beta_{j}=\frac{exp(\mathbf{h}_2^T \mathbf{W}_{2} \hat{\mathbf{u}}_{j})}{\sum_{{j'} \in\mathcal{G}_t} exp(\mathbf{h}_2^T \mathbf{W}_{2} \hat{\mathbf{u}}_{j'})},
\end{equation} 
where $\mathbf{g}^m_t$ denotes the user-level representation of group $g_t$, $\beta_{j}$ indicates the weight of user $u_j$ in the group decision, and $\mathbf{W}_{2}$ along with $\mathbf{h}_2^T$ are learnable parameters. By stacking all the $\mathbf{g}^m_t(t=1,...,K)$, we can get the groups’ user-level representations $\hat{\mathbf{G}}^m \in \mathbb{R}^{2d \times K}$.  

\subsection{Group-Level Preference Network} 

\subsubsection{Co-occurrence Graph Propagation.} 
As mentioned before, merely aggregating user-level representations is insufficient to adequately capture the inherent preferences of the group. Since we only learn the group's inherent preferences independently, we still apply graph neural networks to encode high-order relationships between groups based on group similarity. Specifically, we feed the group embeddings $\mathbf{G} \in \mathbb{R}^{2d \times K}$ to the graph convolutional network, denoted as $\mathbf{G}^{(0)}=\mathbf{G}$. The propagation mechanism at each layer is similar to Equation 1. After passing $\mathbf{G}^{(0)}$ through $L$ convolutional layers, we average the embeddings obtained at each layer to get the final representation $\hat{\mathbf{G}}^t=\frac{1}{L+1}\sum^L_{l=0}\mathbf{G}^{(l)}$.

\subsubsection{Group Gate Aggregation.}
We have obtained the user-level group representation $\hat{\mathbf{G}}^m$ and the group-level representation $\hat{\mathbf{G}}^t$. To adaptively control their combination, we propose a gate aggregator:
\begin{equation}
\hat{\mathbf{G}} = \gamma\hat{\mathbf{G}}^m + (1-\gamma)\hat{\mathbf{G}}^t, 
\end{equation}
\begin{equation} 
\gamma = \sigma(\mathbf{W}^m\mathbf{G}^m + \mathbf{W}^t\mathbf{G}^t + \mathbf{b}^g),
\end{equation} 
where $\mathbf{W}^m$, $\mathbf{W}^t$ and $\mathbf{b}^g$ are learnable parameters and $\sigma$ is the activation function. The $\gamma$ denote the learned weight that balanced the contributions of $\hat{\mathbf{G}}^m$ and $\hat{\mathbf{G}}^t$. By using the gate aggregator, we can obtain the final group representations $\hat{\mathbf{G}}$.   

\subsection{Model Optimization}  
To optimize the DisRec parameters, we unify the loss functions of user, group, and SSL. Firstly, we adopt the pairwise learning task loss function \cite{wang2017item} to optimize user and item representations, which is designed as follows: 
\begin{equation}
\mathcal{L}_{u s e r}=-\sum_{(j, h, h') \in \mathcal{O}} (\hat{r}_{jh}-\hat{r}_{jh'} - 1)^2, \hat{r}_{jh}=\mathbf{u}_j\mathbf{i}_h^T, 
\end{equation} 
where $\mathbf{u}_j$ represents the embedding of user $u_j$, $\mathbf{i}_h$ represents the embedding of item $i_h$. $\mathcal{O}$ represents the training set, in which each one includes the user $u_j$, interacted item $i_h$, and unobserved items $i_h'$. Similarily, the group loss function can be achieved as follows: 
\begin{equation}
\mathcal{L}_{group}=-\sum_{(t, h, h') \in \mathcal{O'}} (\hat{y}_{th}-\hat{y}_{th'} - 1)^2,  \hat{y}_{th}=\mathbf{g}_t\mathbf{i}_h^T, 
\end{equation}  
where $\mathbf{g}_t$ represents the embedding of group $g_t$, $\mathbf{i}_h$ represents the embedding of item $i_h$. $\mathcal{O'}$ represents the training set, in which each one includes the group $g_t$, interacted item $i_h$,  and unobserved items $i_h'$.  

\subsubsection{Contrastive learning.}   
To overcome the long-standing issue of data sparsity in GR, we intend to leverage the rich self-supervised signals from social information to enhance group representations. Specifically, we identify the most important person $u'$ and the least important person $u''$ in group $g_t$ based on Equation 6. By removing $u'$ and $u''$ respectively, we obtain new group representations $g'$ and $g''$. We treat $(g_t, g')$ as a positive example pair and $(g_t, g'')$ as a negative example pair. Our goal is to accurately reflect how changes in group membership impact group decisions. Following S$^2$-HHGR \citep{zhang2021double}, we adopt crossentropy loss to construct the contrastive loss: 
\begin{equation}
\mathcal{L}_{ssl}=-\sum_{t \in \mathcal{G}}\left[\log \sigma\left(f\left(\hat{\mathbf{g}}_t, \hat{\mathbf{g}}_t'\right)\right) +\left[\log \sigma\left(1-f\left(\hat{\mathbf{g}}_t, \hat{\mathbf{g}}_t'' \right)\right)\right]\right],
\end{equation}
where $f\left(\hat{\mathbf{g}}_t, \hat{\mathbf{g}}_t' \right)=\sigma(\hat{\mathbf{g}}_t\mathbf{W}_c\hat{\mathbf{g}}_t'^T)$ is a learnable distance function, and $\hat{\mathbf{g}}_t$ is the representation of group $g_t$. We jointly train $\mathcal{L}_{user}$, $\mathcal{L}_{group}$ and $\mathcal{L}_{ssl}$ using the Adam algorithm. The overall objective is defined as follows: 
\begin{equation}
    \mathcal{L} = \mathcal{L}_{user} + \mathcal{L}_{group} + \delta \mathcal{L}_{ssl}
\end{equation}
where $\delta$ is the hyper-parameter to balance the task of self-supervised learning and supervised learning task for GR. 

\begin{table}[]
\centering
\small
\setlength{\tabcolsep}{1mm}
\begin{tabular}{c|ccccc}
\toprule
Dataset  & \#Users & \#Groups & \#Items & \begin{tabular}[c]{@{}c@{}}\#U-I\\ interactions\end{tabular} & \begin{tabular}[c]{@{}c@{}}\#G-I\\ interactions\end{tabular} \\ \midrule
Mafengwo & 5,275   & 995      & 1,513   & 39,761                                                       & 3,595                                                        \\
Yelp     & 34,504  & 24,103   & 22,611  & 482,273                                                      & 26,883                                                       \\ \bottomrule
\end{tabular}
\caption{Statistics of datasets.}
\label{table1}
\end{table}

\begin{table*}[]
\centering
\small
\setlength{\tabcolsep}{1.5mm} 
\begin{tabular}{@{}c|c|cccccccc|cc@{}}
\toprule
Dataset                   & Metric  & NCF    & SIGR   & HCR    & GroupIM      & S$^2$-HHGR   & SGGCF  & CubeRec & ConsRec      & DisRec           & Improv. \\ \midrule
\multirow{4}{*}{Mafengwo} & HR@5    & 0.3681 & 0.4092 & 0.4824 & 0.5297       & 0.5726 & 0.6241 & 0.6824  & {\underline{0.6925}} & \textbf{0.7266*} & 4.92\%  \\
                          & HR@10   & 0.4365 & 0.4583 & 0.5176 & 0.6040       & 0.5899 & 0.7005 & 0.7699  & {\underline{0.7829}} & \textbf{0.8101*} & 3.47\%  \\
                          & NDCG@5  & 0.2798 & 0.3077 & 0.4557 & 0.4291       & 0.5405 & 0.5069 & 0.5572  & {\underline{0.5675}} & \textbf{0.6045*} & 6.52\%  \\
                          & NDCG@10 & 0.3171 & 0.3347 & 0.4668 & 0.4534       & 0.5473 & 0.5320 & 0.5862  & {\underline{0.5969}} & \textbf{0.6312*} & 5.75\%  \\ \midrule
\multirow{4}{*}{Yelp}     & HR@5    & 0.2622 & 0.3276 & 0.5086 & 0.5875       & 0.5379 & 0.5428 & 0.5553  & {\underline{0.5904}} & \textbf{0.6157*} & 4.29\%  \\
                          & HR@10   & 0.3165 & 0.4274 & 0.5638 & {\underline{0.6687}} & 0.5965 & 0.6074 & 0.6173  & 0.6606       & \textbf{0.7061*} & 5.59\%  \\
                          & NDCG@5  & 0.2890 & 0.2407 & 0.4065 & 0.4992       & 0.4115 & 0.4687 & 0.4602  & {\underline{0.5051}} & \textbf{0.5441*} & 7.72\%  \\
                          & NDCG@10 & 0.3198 & 0.2729 & 0.4241 & 0.5056       & 0.4239 & 0.4867 & 0.4831  & {\underline{0.5154}} & \textbf{0.5646*} & 9.55\%  \\ \bottomrule
\end{tabular}
\caption{Performance comparison of all methods on \textbf{group recommendation} task. The colunm of ``Improv.'' indicates the improvement of the best result of DisRec (boldface) compared with the best baseline (underscore). * indicates the statistical significance for $p$-value \textless{} 0.05 compared with the best baseline.}
\label{group_result}
\end{table*}

\begin{table*}[]
\centering
\small
\setlength{\tabcolsep}{1.5mm} 
\begin{tabular}{c|c|cccccccc|cc}
\toprule
Dataset                   & Metric  & NCF    & SIGR   & HCR          & GroupIM & S$^2$-HHGR & SGGCF  & CubeRec & ConsRec      & \textbf{DisRec}  & Improv. \\ \midrule
\multirow{4}{*}{Mafengwo} & HR@5    & 0.4759 & 0.4623 & {\underline{0.5804}} & 0.0969  & 0.4734 & 0.4513 & 0.1750  & 0.5736       & \textbf{0.6608*} & 13.85\% \\
                          & HR@10   & 0.5603 & 0.5569 & 0.6186       & 0.1323  & 0.5580 & 0.5286 & 0.3002  & {\underline{0.6346}} & \textbf{0.7020*} & 10.62\% \\
                          & NDCG@5  & 0.4447 & 0.4273 & {\underline{0.5456}} & 0.0686  & 0.3834 & 0.3042 & 0.1045  & 0.5436       & \textbf{0.5795*} & 6.21\%  \\
                          & NDCG@10 & 0.4702 & 0.4506 & 0.5775       & 0.0801  & 0.4110 & 0.3370 & 0.1442  & {\underline{0.5829}} & \textbf{0.6115*} & 4.91\%  \\ \midrule
\multirow{4}{*}{Yelp}     & HR@5    & 0.4029 & 0.4153 & 0.4774       & 0.0998  & 0.4465 & 0.4310 & 0.1611  & {\underline{0.4903}} & \textbf{0.5668*} & 15.60\% \\
                          & HR@10   & 0.4991 & 0.5286 & {\underline{0.5499}} & 0.1560  & 0.5168 & 0.5138 & 0.1957  & 0.5319       & \textbf{0.6245*} & 13.57\% \\
                          & NDCG@5  & 0.3981 & 0.3042 & 0.4375       & 0.0652  & 0.3841 & 0.3253 & 0.1259  & {\underline{0.4640}} & \textbf{0.5028*} & 8.36\%  \\
                          & NDCG@10 & 0.4516 & 0.3370 & 0.4870       & 0.0865  & 0.4772 & 0.3426 & 0.1371  & {\underline{0.5083}} & \textbf{0.5468*} & 7.57\%  \\ \bottomrule
\end{tabular}
\caption{Performance comparison of all methods on \textbf{user recommendation} task.}
\label{user_result}
\end{table*}

\section{Experiment}
\subsection{Experimental Settings}
\subsubsection{Datasets.}
We conduct experiments on two real-world public datasets: Mafengwo, published by \citep{cao2018attentive}, and Yelp, published by \citep{yin2019social}. Mafengwo is a tourism website where users create or join group travel activities. Yelp is a typical check-in dataset for various restaurants in the US. Table 1 reports the statistics of these datasets.  

\subsubsection{Evaluation Metrics.}
To evaluate the performance of the Top-K recommendation for all methods, we employ the widely adopted metrics NDCG@K and HR@K with K=$\{5, 10\}$, as in \cite{wu2023consrec}. All items are considered as candidates for fairness. The permutation test \citep{nichols2002nonparametric} is adopted for significance tests. 

\subsubsection{Baselines.} 
To evaluate the effectiveness of DisRec, we compared it with the following baselines: \textbf{NCF} \citep{he2017neural} treats a group as a virtual user and ignores the member information of the group. \textbf{SIGR} \citep{yin2019social} integrates users’ global and local social information to improve GR. \textbf{GroupIM} \citep{sankar2020groupim} maximizes mutual information between users and groups to overcome the group-level data sparsity. \textbf{HCR} \citep{jia2021hypergraph} proposes a dual channel hypergraph convolutional network to capture member-level and group-level preferences. \textbf{S$^2$-HHGR} \citep{zhang2021double} integrates hierarchical hypergraph learning and self-supervised learning for improved group representation. \textbf{CubeRec} \citep{chen2022thinking} utilizes a hypercube vector space to replace point embeddings for representing group preferences. \textbf{SGGCF} \citep{Li2023SelfSupervisedGG} explores two kinds of contrastive learning module to capture the implicit relations between groups and items. \textbf{ConsRec} \citep{wu2023consrec} designs three views to capture the consensus information within groups. 


\subsubsection{Implementation Details.} 
For the general settings, the embedding size is set to 64, the batch size is 512, and the number of negative samples is 10. For the baseline models, we refer to their best parameter setups reported in the original papers. For our model, we set the convolutional layer $L$ to 3, the SSL weight $\delta$ to 0.5 and tune the learing rate in $[1e-4, 1e-3]$, and the dropout rate in $[0.1, 0.5]$.

\subsection{Overall Performance}
The experimental performance on GR task and user recommendation (UR) task are shown in Table 2 and Table 3, respectively. We observed that in both GR and UR tasks, our proposed DisRec consistently outperforms all baselines on two datasets in terms of HR@K and NDCG@K metrics.

For the GR task, our model gains the greatest advantage over single-strategy methods that use only hypergraphs or SSL (HCR, GroupIM), with average increases of 30.4$\%$ in H@10 and 29.8$\%$ in N@10 across two datasets. Compared to mixed-strategy methods that integrate both (hyper) graph and SSL (S$^2$-HHGR, SGGCF), our improvements are slightly smaller but still significant, with average increases of 21.9$\%$ in H@10 and 20.8$\%$ in N@10 across two datasets. This is because DisRec not only designs superior social hypergraphs and more effective social-based SSL, but also overcomes the preference bias issue of previous models by disentangling user preferences and social influence. Additionally, compared to models that also use social information (SIGR and CubeRec), our model achieves improvements of at least 
12.3$\%$ in N@10. This improvement may be attributed to our social hypergraph's ability to better capture users' social influence both within and outside their groups.

For the UR task, our model also demonstrates notable advantages. For instance, compared to the strongest baseline model, ConsRec, our model achieve an average improvement of 12.10$\%$ in H@10 and 6.24$\%$ in N@10 across two datasets. We attribute this to disentangled user representations, which help mitigate preference bias in GR as well as enhance the performance of the UR task. 

\begin{table}[]
\centering
\small
\setlength{\tabcolsep}{1.5mm} 
\begin{tabular}{@{}cccccc@{}}
\toprule
Dataset                                        & Metric  & w/o. S & w/o. P & w/o. SSL                          & \textbf{DisRec}  \\ \midrule
\multicolumn{1}{c|}{\multirow{2}{*}{Mafengwo}} & HR@10   & 0.7698 & 0.7888 & \multicolumn{1}{c|}{{\underline{0.8050}}} & \textbf{0.8101*} \\
\multicolumn{1}{c|}{}                          & NDCG@10 & 0.5916 & 0.6125 & \multicolumn{1}{c|}{{\underline{0.6214}}} & \textbf{0.6312*} \\ \midrule
\multicolumn{1}{c|}{\multirow{2}{*}{Yelp}}     & HR@10   & 0.6625 & 0.6869 & \multicolumn{1}{c|}{{\underline{0.6942}}} & \textbf{0.7061*} \\
\multicolumn{1}{c|}{}                          & NDCG@10 & 0.5267 & 0.5423 & \multicolumn{1}{c|}{\underline{0.5514}}              & \textbf{0.5646*} \\ \bottomrule
\end{tabular}
\caption{The comparison between DisRec and its variants.}
\label{tab:my-table}
\end{table}

\subsection{Ablation Study} 
We adopt two sets of embeddings to disentangle user preferences and social influence and employ a social-based SSL strategy to enhance group representations. To verify the effectiveness of each main components of DisRec, we conduct the ablation study to analyze their contribution. Due to space limitations, Table 4 only shows the results on H@10 and N@10 for two datasets, where ``w/o. S'', ``w/o. P'', and ``w/o. SSL'' represent the variants of removing social influence embeddings, preferences embeddings, and the SSL strategy, respectively. From Table 4, we observe that removing any part results in a performance degradation across all datasets. For example, on the Yelp dataset, ``w/o. SSL'' shows a $2.3\%$ decrease in NDCG@10 compared to DisRec. Furthermore, we observe that the variant ``w/o. SSL'' performs the best among all variants, indicating that disentangled user embeddings have a greater impact on the model performance. However, the variant ``w/o. S'' outperforms the variant ``w/o. P'', suggesting that solely modeling user preferences is insufficient and further emphasizing the importance of disentangled user contributions in GR.

\begin{figure}[t]
  \centering
  \begin{subfigure}{0.22\textwidth}
    \subcaptionbox*{$L-$NDCG@10}{\includegraphics[width=\textwidth]{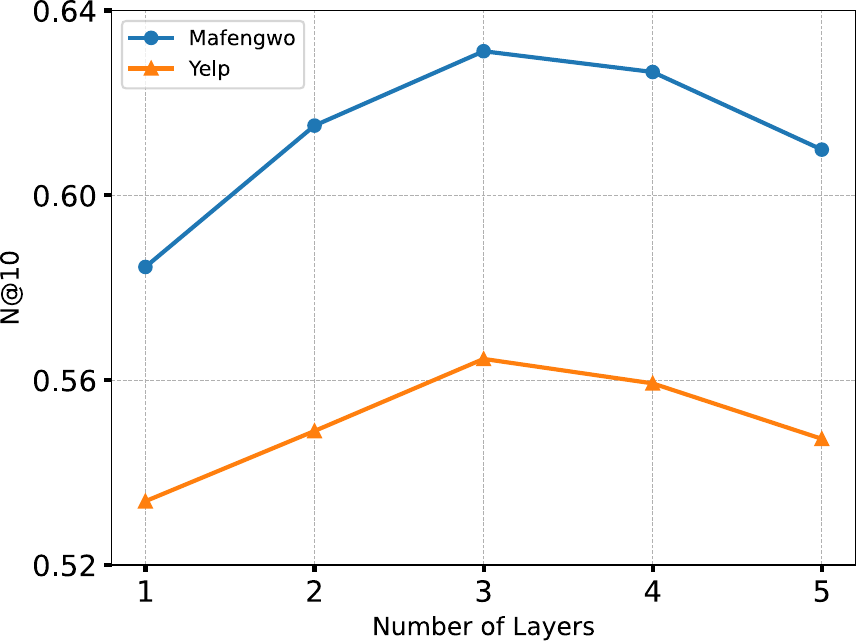}}
  \end{subfigure}
  \quad
  \begin{subfigure}{0.22\textwidth}
    \subcaptionbox*{$L-$HR@10}{\includegraphics[width=\textwidth]{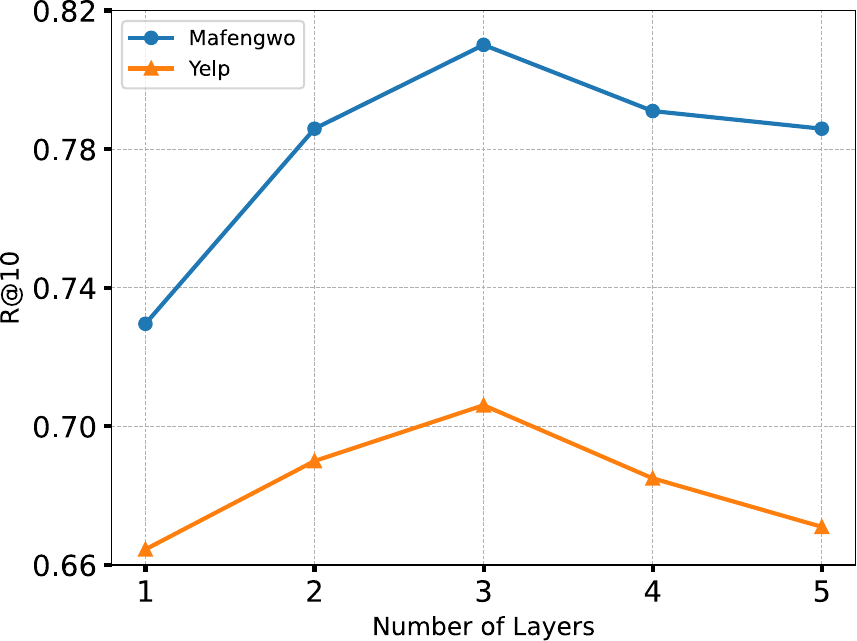}}
  \end{subfigure}
  \begin{subfigure}{0.22\textwidth}
    \subcaptionbox*{$\delta-$NDCG@10}{\includegraphics[width=\textwidth]{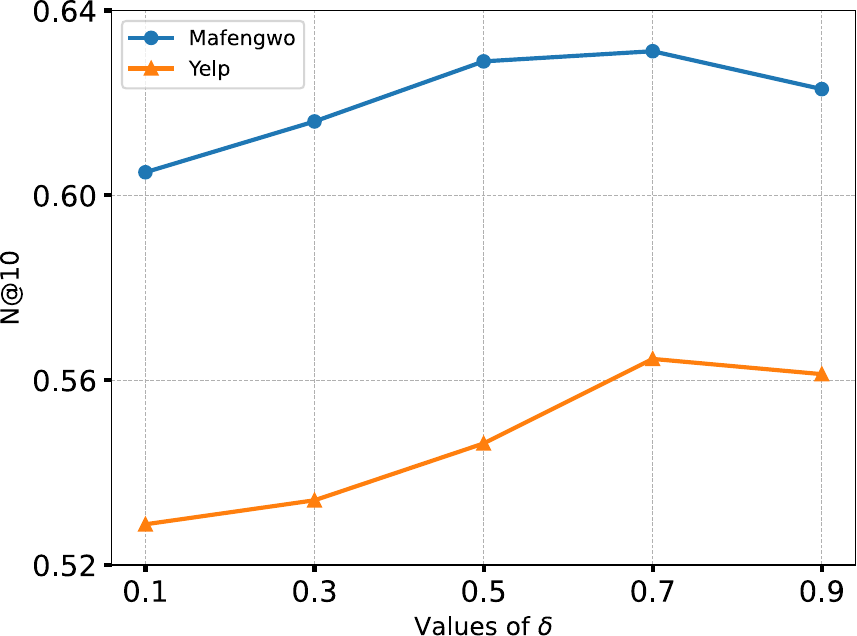}}
  \end{subfigure}
  \quad
  \begin{subfigure}{0.22\textwidth}
    \subcaptionbox*{$\delta-$HR@10}{\includegraphics[width=\textwidth]{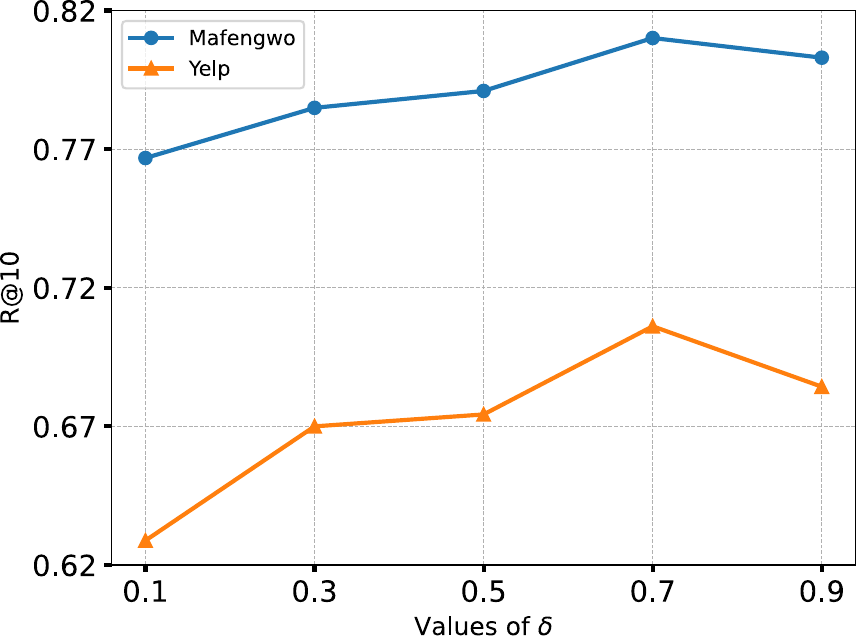}}
  \end{subfigure}
  \caption{The influence of different model hyperparameters.}
  \label{figure3}
\end{figure}

\subsection{Hyperparameter Analysis} 
To investigate the impact of important hyperparameters in DisRec, we evaluate the number of graph convolution layers $L$ and the balance coefficient of SSL task $\delta$ on two datasets. We adjust each hyperparameter while others remain unchanged, and the results are reported in Figure 4. 

\textbf{Impact of $L$}. The performance of graph convolutional networks is influenced by the number of network layers $L$. We tune $L$ in $\{1, 2, 3, 4, 5\}$. In general, DisRec benefits a larger $L$ across all datasets, achieving the best performance with $L=3$. Performance declines with excessive layers (e.g., $L=5$) due to over-smoothing, while too few layers fail to capture complex higher-order relationships. 

\textbf{Impact of $\delta$}. The coefficient $\delta$ in Eq. (12) controls the SSL regularization effect. To analyze the impact of $\delta$ on our model, the coefficient $\delta$ is set among $\left\{0.1,0.3,0.5,0.7,0.9\right\}$. DisRec's performance increases as $\delta$ rises, until $\delta$ reaches 0.7, after which it decreases with further increases in $\delta$. The worst performance occurs when $\delta=0.1$, suggesting that utilizing our proposed social-based SSL strategy is necessary to address the issue of group data sparsity.
However, excessively high $\delta$ values cause the model to overly focus on high-weight individuals, resulting in suboptimal outcomes. 

\begin{figure}[t]
    \centering
    \begin{subfigure}[b]{0.42\textwidth}
        \flushleft
        \includegraphics[width=\textwidth]{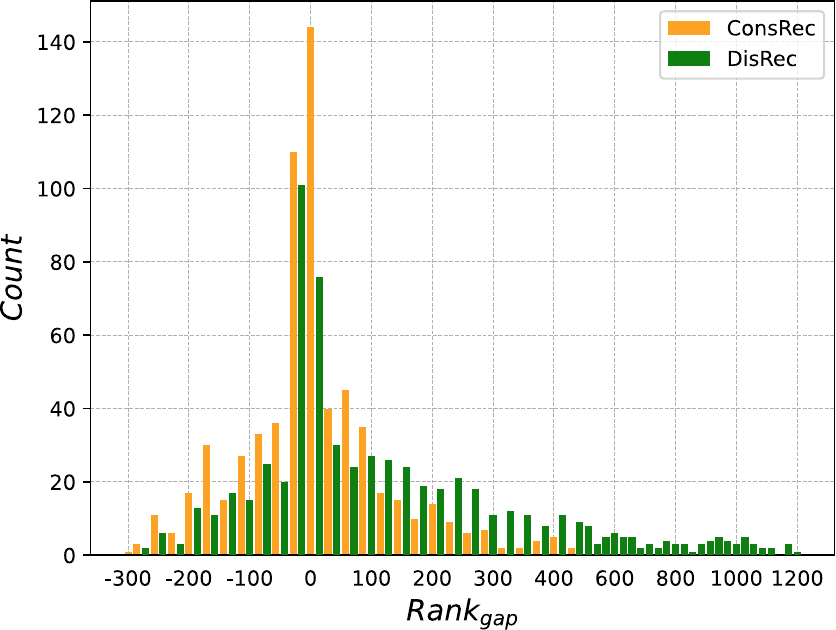}
        \caption{Distribution of $Rank_{gap}$ values.}
        \label{fig:subfig1}
    \end{subfigure}
    \hfill
    \begin{subfigure}[b]{0.35\textwidth}
        \hfill
        \includegraphics[width=\textwidth]{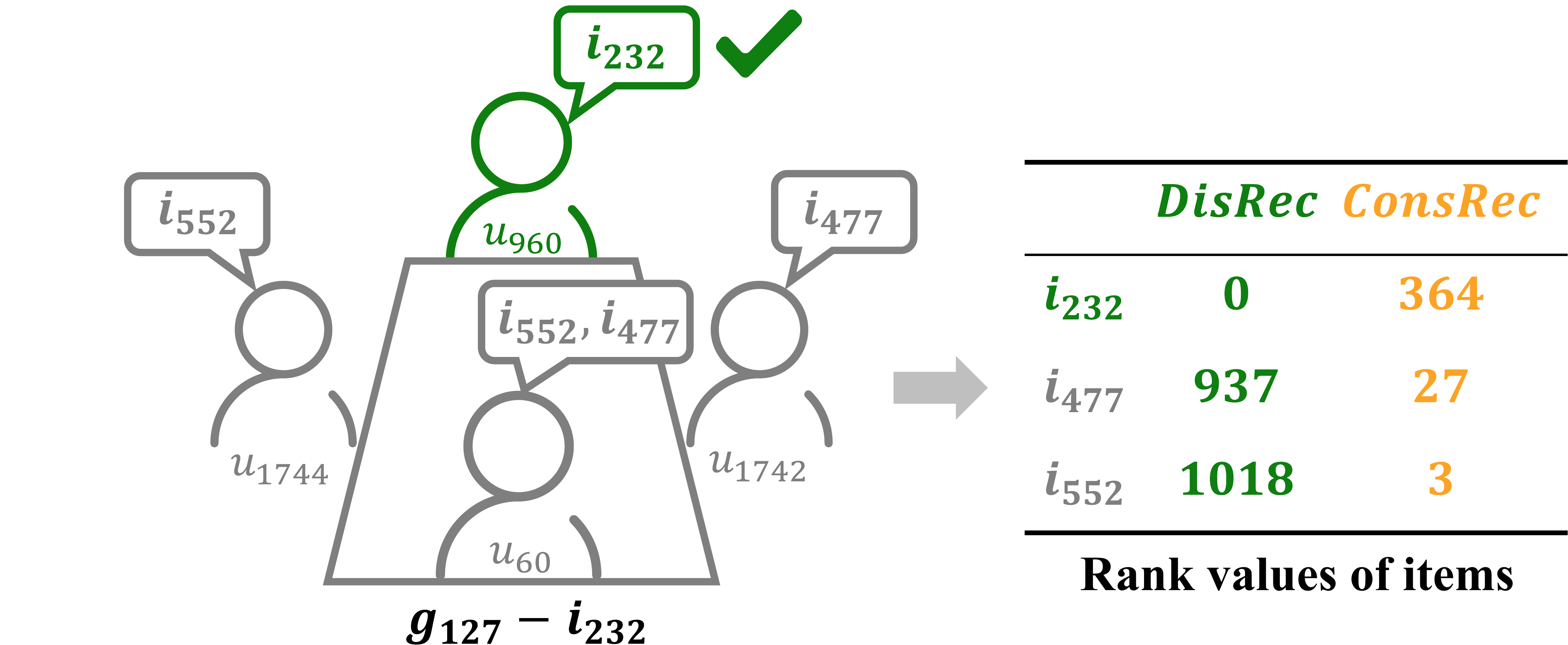}
        \caption{A visualized example from group $g_{127}$.}
        \label{fig:subfig2}
    \end{subfigure}
    \caption{Results of DisRec evaluation on synthetic data.}
    \label{fig:mainfig}
\end{figure} 

\subsection{Case Study} 
To further investigate the effectiveness of DisRec in addressing preference bias in GR, we conduct a case study using a synthetic dataset inspired by \citep{wang2021clicks}. Specifically, we generate two fake items for each positive group-item pair in the test set of the Mafengwo dataset, with these fake items interacted with by at least 2/3 of the group members. Models with preference bias often favor items aligning with the majority preferences, even if these are not the group's final choices. We define $Rank_{gap} = Rank_{fake} - Rank_{true}$ to measure the model's ability to overcome preference bias, where $Rank_{fake}$ and $Rank_{true}$ represent the ranks of the paired fake and real items, respectively. A model that effectively mitigates preference bias should have a high $Rank_{gap}$. Figure 5(a) presents the $Rank_{gap}$ values for DisRec and the strongest baseline model, ConsRec, across 719 test samples. It is evident from the figure that DisRec has a lower $Count$ value near a $Rank_{gap}$ of 0 compared to ConsRec. The maximum $Rank$ value for ConsRec peaks around 400, while DisRec's $Rank$ values are more evenly distributed between 400 and 1200. This is attributed to DisRec's successful disentangling of user preferences and social influence causes. Additionally, Figure 5(b) visualizes a randomly selected test case $g_{127}-i_{232}$. Here, group $g_{127}$ interacts with item $i_{232}$, while items $i_{477}$ and $i_{552}$, which are popular among most members, are set as fake items. The right table shows that DisRec correctly assigns a higher rank to item $i_{232}$, whereas ConsRec is misled by the fake items, demonstrating DisRec's effectiveness in mitigating preference bias. 

\section{Conclusion}
In this paper, we address the issue of preference bias in previous GR models and propose an innovative model DisRec, which solves this problem by disentangling users' preferences and social influence. Additionally, DisRec adopts a social-based SSL strategy to alleviate the issue of group data sparsity by generating higher quality positive and negative samples to regularize group representations. Through extensive experiments on two widely used datasets, we demonstrate the superiority of our model in both GR and UR tasks. Furthermore, a case study focused on the issue of preference bias confirms DisRec's ability to distinguish between real and fake items.

\section{Acknowledgments}
This work is funded by STI 2030-Major Projects 2021ZD0200500, National Natural Science Foundation of China (under project No. 62377013), and the Fundamental Research Funds for the Central Universities. It is also supported by Natural Science Foundation of Shanghai, China (under project No. 22ZR1419000), the Research Project of Changning District Science and Technology Committee (under project No. CNKW2022Y37), and the Medical Master’s and Doctoral Innovation Talent Base Project of Changning District (under project No. RCJD2022S07). 

\bibliography{aaai25}

\end{document}